\documentclass[]{111}

\usepackage{epstopdf}
\usepackage[caption=false]{subfig}

\usepackage{url}
\usepackage{hyperref}
\usepackage{float}

\usepackage[natbibapa]{apacite}
\usepackage{url}
\usepackage{float}
\AtBeginDocument{%
}

\usepackage[T1]{fontenc}
\usepackage{lmodern}
\usepackage{booktabs}
\usepackage{multirow}
\usepackage{tabularx}
\usepackage{threeparttable}

\theoremstyle{plain}

\theoremstyle{definition}

\theoremstyle{remark}

\begin{document}


\title{Exploring Sidewalk Sheds in New York City through Chatbot Surveys and Human–Computer Interaction}

\author{
\name{Junyi Li \textsuperscript{a,b}, Zhaoxi Zhang $^\ast$  \textsuperscript{a,d}\thanks{$^\ast$ Corresponding Author: Zhaoxi Zhang, Department of Urban and Regional Planning, University of Florida. Email: zhang.zhaoxi@ufl.edu}, Tamir Mendel \textsuperscript{c,e} and Takahiro Yabe \textsuperscript{a,c}} 
\affil{
\textsuperscript{a}Center for Urban Science + Progress, Tandon School of Engineering, New York University, Brooklyn, 11201, USA;
\textsuperscript{b}Department of Civil and Urban Engineering, Tandon School of Engineering, New York University, Brooklyn, 11201, USA;
\textsuperscript{c}Department of Technology Management and Innovation, Tandon School of Engineering, New York University, Brooklyn, 11201, United States of America;
\textsuperscript{d}Department of Urban and Regional Planning, College of Design, Construction and Planning, University of Florida;
\textsuperscript{e}School of Information Systems, The Academic College of Tel Aviv-Yaffo, Tel Aviv, Israel;
}
}

\maketitle

\begin{abstract}
Sidewalk sheds are a common feature of the streetscape in New York City, reflecting ongoing construction and maintenance activities. However, policymakers and local business owners have raised concerns about reduced storefront visibility and altered pedestrian navigation. 
Although sidewalk sheds are widely used for safety, their effects on pedestrian visibility and movement are not directly measured in current planning practices. To address this, we developed an AI-based chatbot survey that collects image-based annotations and route choices from pedestrians, linking these responses to specific shed design features, including clearance height, post spacing, and color.
This AI chatbot survey integrates a large language model (e.g., Google’s \textit{Gemini-1.5-flash-001} model) with an image-annotation interface, allowing users to interact with street images, mark visual elements, and provide structured feedback through guided dialogue.
To explore pedestrian perceptions and behaviors, this paper conducts a grid-based analysis of entrance annotations and applies logistic mixed-effects modeling to assess sidewalk choice patterns. Analysis of the dataset ($n=25$) shows that: (1) the presence of scaffolding significantly reduces pedestrians’ ability to identify ground-floor retail entrances, and (2) variations in weather conditions and shed design features significantly influence sidewalk selection behavior. By integrating generative AI into urban research, this study demonstrates a novel method for evaluating sidewalk shed designs and provides empirical evidence to support adjustments to shed guidelines that improve the pedestrian experience without compromising safety.
\end{abstract}


\section{Introduction}

Sidewalk sheds, commonly known as scaffolding, have become an almost permanent feature of New York City’s streetscape.
Originally intended as temporary protective structures mandated under Local Law 11 \citep{NYC_Local_Law_11_1998} for periodic façade inspections and maintenance, many now remain in place for extended periods due to construction backlogs, ownership incentives, and regulatory delays.
As of 2024, more than 8,500 sidewalk sheds cover approximately 330 miles of city streets, with an average installation duration of nearly 500 days; 3,762 of these structures have persisted for over a year \citep{NYCDOBmap2025}. Their widespread and prolonged presence has raised concerns about the cumulative effects of sidewalk sheds on urban design quality, public realm experience, and everyday pedestrian life.

A growing body of research and media reports suggests that sidewalk sheds alter the visual and spatial characteristics of sidewalks. These scaffolding structures may diminish storefront visibility and visual permeability \citep{Menkes2003,kang2016high} by obstructing sightlines, reducing natural light, and narrowing walking corridors. These changes can complicate navigation and spatial decision-making \citep{Numina2023}, potentially depressing commercial activity \citep{Mayor2024}. 
From an urban design perspective, sidewalk sheds can over-enclose the pedestrian environment while reducing visual permeability and legibility, disrupting key spatial characteristics that support walkability, way-finding, and perceived safety \citep{EwingHandy2016,Gehl2010,Lynch1960}.
In response to mounting concerns, initiatives such as New York City’s ‘Get Sheds Down’ campaign \citep{NYC_SidewalkSheds_2025} aim to accelerate removals and promote aesthetic upgrades. Yet, despite increasing policy attention, there remains limited empirical evidence on how specific design features of sidewalk sheds—such as color, height, and post spacing—along with contextual factors, such as weather, shape street-level pedestrian behavior.

The emergence of human-computer interaction and conversational AI in urban planning creates opportunities to scale participation and enable flexible data collection, effectively integrating public feedback into policymaking processes \citep{EvansCowley2010,portela2021interfacing}.
Many researchers have emphasized the benefits of digital tools for participatory urban research; however, these tools have not been widely adopted in practice \citep{Cantijoch2019,Panagiotopoulou2017,Hasan2024}. 
To address these gaps, this study develops a novel, generative-AI–driven chatbot survey that engages participants in structured annotation and dialogue tasks using real-world images of sidewalk shed conditions. The platform combines quantitative measures (e.g., storefront entrance recognition) with qualitative explanations for navigation preferences collected through adaptive follow-up questioning. 
This study aims to enhance the understanding of how sidewalk shed design affects pedestrian experiences, with the goal of supporting regulations and design practices that balance safety with street-level accessibility and urban vitality.

\section{Related Work}
\subsection{Sidewalk Sheds and Pedestrian Behavior}

Sidewalk sheds are temporary structures built to protect people and property, serving essential safety functions. However, researchers have described the negative impact of sidewalk sheds on pedestrian behavior. Sidewalk sheds hinder the flow of pedestrian traffic, leading individuals to adjust their routes. A study by \cite{Numina2023} revealed that pedestrians were 53\% more likely to walk in the street when scaffolding obstructed sidewalks, increasing the risk of accidents. The presence of sidewalk sheds also obstructs pedestrian access to storefronts \citep{Menkes2003}. Obscured visibility of a storefront may significantly reduce a store's attractiveness to pedestrians and its ability to capture foot traffic \citep{kang2016high}. Research indicates that businesses located in buildings with sidewalk sheds experience a monthly decrease in spending by Mastercard cardholders ranging from $3,900$ to $9,500$ \citep{Mayor2024}. 

While sidewalk sheds have been associated with negative impacts on pedestrian behavior and retail visibility, they also offer important benefits under adverse weather conditions. By providing shelter during rain, sidewalk sheds help sustain pedestrian movement along the street. Similarly, during periods of high temperatures, sidewalk sheds provide shaded environments, making them preferable routes for pedestrians seeking to avoid direct solar exposure. Research has shown that pedestrians actively select shaded routes in urban areas during hot weather, highlighting the role of sidewalk structures in enhancing thermal comfort \citep{watanabe2016effect}. Additionally, \cite{Aneja2020} proposes transforming New York City’s sidewalk sheds into modular structures that provide weather protection. Thus, despite their drawbacks, sidewalk sheds can mitigate environmental discomfort and support continuous pedestrian movement under challenging weather conditions.

A range of plans have been proposed to mitigate the negative impacts by modifying the appearance and structure of sidewalk sheds. For example, Mayor Adams and the City Council proposed reforms to sidewalk shed design standards, including new color options, improved lighting, higher ceilings, and the elimination of cross-bracing for sheds in parks \citep{CitizensUnion2023}. Among these features, the height and post spacing of sidewalk sheds are particularly critical, as greater clearance allows for increased natural light penetration and improved storefront visibility, thereby alleviating some adverse impacts on pedestrian experience and commercial activity \citep{Menkes2003}.

While the relationship between sidewalk sheds and pedestrian behavior is increasingly recognized, existing studies provide limited quantitative research and lack empirical assessments of how design interventions may mitigate adverse effects, as well as how weather variability may moderate these impacts. Quantifying the extent of these impacts is crucial for informing policy decisions regarding the allowable duration of sidewalk shed installations, balancing the needs of urban development with the pedestrian experience. Moreover, identifying the influences of specific design features can inform regulatory standards to enhance public safety and urban quality. Given these significant yet understudied impacts on pedestrian experience and streetscape visibility, effective public engagement has become critical for guiding design improvements, with digital technologies increasingly enabling broader and more structured participatory processes.

\subsection{Human-computer Interaction in Citizen Science}

Citizen science broadly refers to the involvement of non-experts in scientific data collection, analysis, and interpretation, and has gained substantial momentum with the proliferation of smartphones and networked information and communication technologies (ICT) \citep{preece2016citizen}. 
Traditional citizen science and participatory planning studies frequently rely on surveys or open-ended feedback mechanisms that collect participants’ reflections after the fact \citep{Arnstein1969,Innes2004}. As noted in prior work, unstructured or weakly contextualized input often leads to inconsistently formatted data and limited spatial specificity, thereby constraining its analytical and policy relevance \citep{Kahila2019,AlKodmany1999}. 

Addressing the challenge of uninformative citizen science methods, digital technologies have accelerated the shift to e-participatory approaches in urban planning. Digital platforms enable more widespread and open-ended public input for civic engagement and public participation. Grassroots digital platforms, such as Decidim \footnote{\url{https://decidim.org/}}, CityScope \footnote{\url{https://cityscope.media.mit.edu/}}, and Ushahidi \footnote{\url{https://www.ushahidi.com/}}, provide bottom-up channels for civic expression and have been shown to support ongoing dialogue between planners and the public, shifting the nature of engagement from passive input to active co-production \citep{Gordon2011_Planning}. 

While digital tools can dramatically scale participation, the design of human-computer interaction shapes the types of human knowledge, perception, and behavior that can be captured. While large language models enable interactive dialogue \citep{mctear2024transforming}, recent research positions conversational interfaces as a human–computer interaction (\textit{HCI}) mechanism that transforms citizen participation from passive reporting to interactive, task-guided engagement. This approach enables structured yet flexible data collection while lowering participation barriers \citep{portela2021interfacing}. More broadly, work on human-centered and human–AI collaboration emphasizes that AI in citizen science should not merely automate analysis but also actively shape task presentation and elicit human input, positioning citizen science as a testbed for human–AI collaboration in perceptual and context-dependent tasks \citep{rafner2022mapping}.

Although recent advances in human–computer interaction and citizen science have introduced a range of digital tools for participatory urban research, empirical applications that systematically measure situated perceptions and behaviors remain limited~\citep{Cantijoch2019,Panagiotopoulou2017}. While interactive platforms have expanded opportunities for engagement, their potential to support task-based, context-aware measurement in real urban environments has not been fully explored. This study addresses this gap by leveraging conversational AI as a human-computer interaction approach to elicit structured street-level perceptual data, demonstrating how interactive dialogue can generate actionable evidence for participatory urban planning

\subsection{Conversational AI in Urban Management}

Conversational AI is a dialogue-based interaction paradigm that enables users to engage with digital systems through natural language and has been increasingly adopted across application domains, including healthcare \citep{Nadarzynski2024}, education \citep{Graesser2005}, and urban planning \citep{Wolniak2024}. These systems are designed to facilitate human-like interactions, enabling more effective communication between users and AI-powered agents. One significant application of conversational AI is information elicitation, in which AI agents conduct interviews, collect qualitative data, and assist in decision-making processes via structured protocols \citep{Bickmore2016}.

Current studies have investigated various strategies to enhance the capabilities of conversational agents (\textit{CAs}) in eliciting information. \cite{Hu2024} developed a framework that enables \textit{CAs} to dynamically generate follow-up questions, thereby improving user engagement and response quality. Similarly, \cite{Ruan2019} proposed an adaptive questioning approach to refine \textit{CAs}' abilities in gathering detailed insights based on user input. By leveraging natural language processing (\textit{NLP}) techniques, conversational AI systems continue to evolve, addressing challenges such as bias, interpretability, and contextual understanding \citep{Jentzsch2019}. Future research should focus on improving these capabilities to ensure that conversational AI systems become more reliable, effective, and user-friendly in diverse applications, including urban governance and policymaking.

In urban planning and management, recent studies have also investigated how AI chatbots provide a unique opportunity to integrate public feedback into policymaking \citep{Hasan2024} and have demonstrated that AI-driven conversational agents can streamline participatory urban design by collecting citizens’ opinions and synthesizing real-time insights \citep{Luusua2023,Elgohary2024,Bosco2024}. For example, AI chatbots have been utilized in disaster management to facilitate the collection of critical situational awareness during emergencies, highlighting their potential in high-stakes environments \citep{Ye2021}. Additionally, \cite{ferrara2019ai} studies demonstrate that phone-based AI chatbots outperform traditional survey and participatory tools by enabling real-time, context-aware collection of citizen feedback, achieving higher response efficiency, improving data quality, and accelerating the synthesis of actionable insights in urban planning, participatory design, and emergency management settings 

However, AI chatbots are constrained by their inability to maintain conversational continuity, process spatial information, fully grasp contextual subtleties, ensure consistent accuracy, and overcome significant development and maintenance costs \citep{Hasan2024, laymouna2024roles, lin2023review}. It is crucial to address these limitations through continuous improvement, the integration of contextual understanding, the mitigation of biases, and the implementation of cost-effective strategies.

\subsection{Study Goals}

Although sidewalk sheds are widely recognized for their influence on pedestrian behavior and commercial activity, existing research remains qualitative and lacks empirical validation of design interventions. Taking advantage of recent advances in human–computer interaction and generative AI, this study proposes an image-based interactive survey method that combines visual annotation with conversational prompts to capture pedestrian perceptions and navigation decisions. 
To examine the feasibility of using HCI and AI chatbots as a method for understanding the impact of sidewalk sheds on human perception,  this study aims to examine: 

\begin{enumerate}
    \item To what extent do sidewalk sheds impede access to storefront entrances and reduce storefront visibility and legibility?
    \item How do sidewalk sheds influence pedestrians’ navigation preferences, including their choice of which side of the sidewalk to walk on?
    \item How do variations in sidewalk shed designs and weather conditions affect pedestrian perceptions and behaviors?
\end{enumerate}

Overall, this paper aims to contribute empirical evidence to inform the refinement of urban design standards and policymaking strategies related to sidewalk sheds that balance structural safety with pedestrian accessibility and economic vitality.

\section{Methods}

\subsection{Features and Site Selection}
In New York City, the widespread deployment of sidewalk sheds presents a significant challenge to the streetscape and pedestrian environment. Under Local Law 11, sidewalk sheds are required during regular facade inspections for buildings over six stories, which occur every five years \citep{Chaudhuri2015}. However, delays in construction or owner noncompliance may allow these structures to remain for extended periods, sometimes indefinitely. As of 2023, over $8,500$ sidewalk sheds covered approximately $330$ miles of city streets, with an average duration of $498$ days per installation \citep{Fontan2023}. The prolonged presence of these structures can degrade neighborhood pedestrian environments, obstruct pedestrian flow, and reduce storefront visibility and access, making New York City a compelling empirical setting for studying the effects of sidewalk shed design.

In response to these concerns, the 2022 New York City Building Code (Chapter 33) introduced targeted revisions intended to reduce the physical and visual impacts of sidewalk sheds \citep{NYCDOB2022}. The revised code focuses on three design features that directly influence the pedestrian experience. First, the minimum clearance height was increased from $8$ to $12$ feet to enhance natural light penetration and reduce enclosure effects. Second, color regulations were expanded to allow metallic gray, white, or hues matching building facades. Under earlier regulations, sidewalk sheds were required to be painted a uniform hunter green, regardless of architectural context, which limited their visual compatibility with surrounding facades. Third, updated structural guidelines require a minimum horizontal span of $10$ feet between vertical posts, with cross bracing positioned no lower than $8$ feet above the sidewalk to minimize obstructions to pedestrian flow. Together, these revisions identify color, clearance height, and post spacing as the primary policy-relevant dimensions through which sidewalk shed design can be evaluated.

\begin{table}[ht]
\centering
\caption{\textbf{Physical characteristics of sidewalk sheds across study sites.}
Clearance height is classified as High if it is greater than or equal to $11.5$ feet, and as Low if it is less than or equal to $9.5$ feet.
Post spacing is classified as Small if the distance between vertical posts is less than or equal to $8$ feet, and as Large if it is greater than or equal to $9$ feet.}
\label{tab:site}
\begin{tabularx}{\textwidth}{l *{6}{>{\centering\arraybackslash}X}}
\toprule
\multirow{2}{*}{\textbf{Feature}} &
\multicolumn{2}{c}{\textbf{Location A}} &
\multicolumn{2}{c}{\textbf{Location B}} &
\multicolumn{2}{c}{\textbf{Location C}} \\
\cmidrule(lr){2-7}
 & \textbf{Current} & \textbf{Past}
 & \textbf{Current} & \textbf{Past}
 & \textbf{Current} & \textbf{Past} \\
\midrule
Color & Green & Green & White & Blue & Mint & Green \\
Clearance Height & 11.5 ft & 8.5 ft & 12 ft & 9 ft & 12 ft & 8.5 ft \\
Post Spacing & 7.5 ft & 9.5 ft & 8 ft & 10.5 ft & 8 ft & 9.5 ft \\
\bottomrule
\end{tabularx}

\begin{tablenotes}[flushleft]
\footnotesize
\item \textit{Note:} \textit{Green} indicates the traditional “hunter green” required under pre-2022 regulations.
\textit{White} indicates one of the newly permitted high-luminance neutral tones under the 2022 code revision. 
\textit{Mint} denotes a light green variant intended to soften the visual impact while maintaining contextual compatibility with façade tones. 
\textit{Blue} represents a non-traditional color used in earlier installations that does not align with current recommended façade-matching or neutral color guidance
\end{tablenotes}

\end{table}

\begin{figure}
    \centering
    \includegraphics[width=1\linewidth]{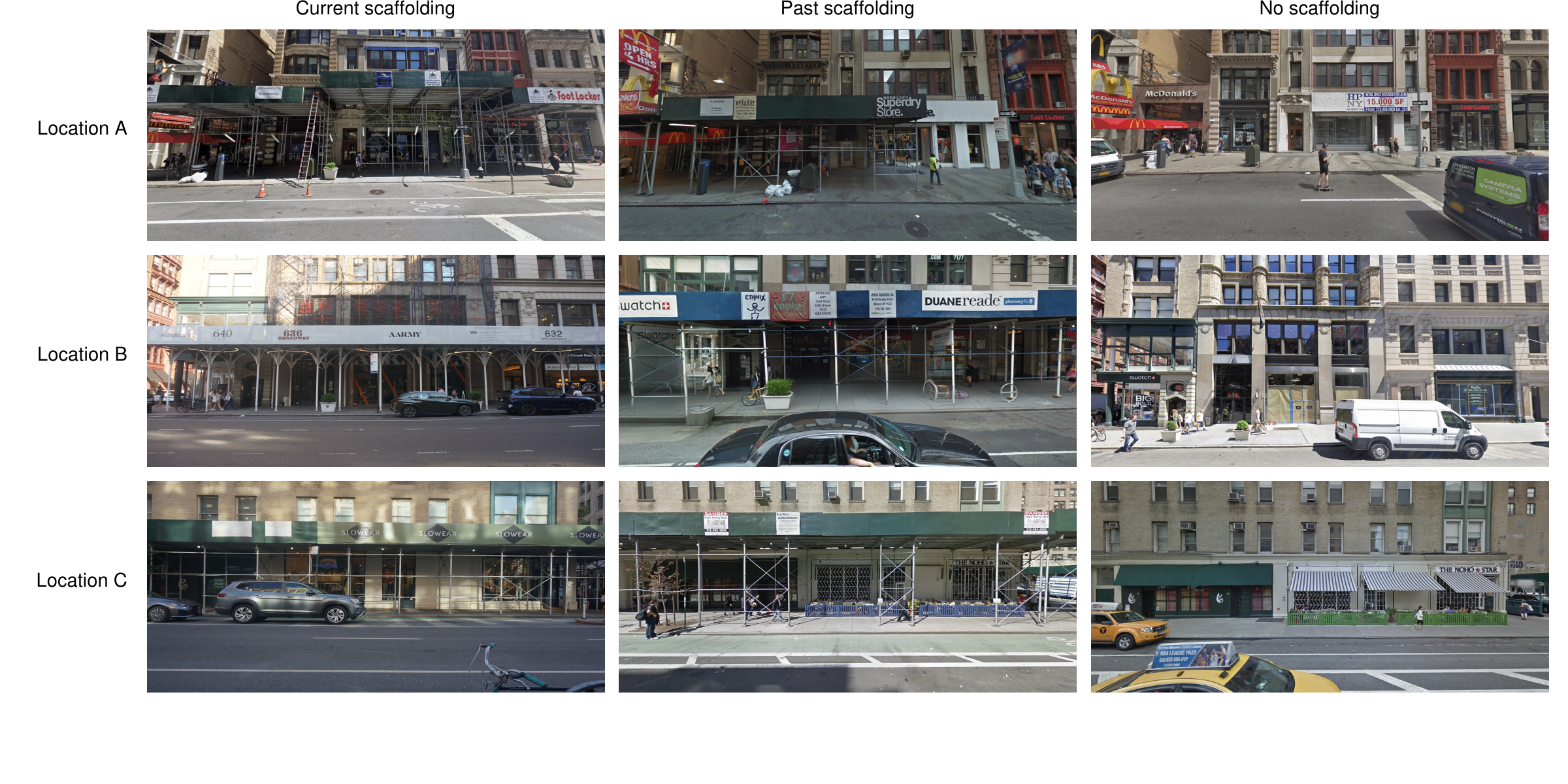}
    \caption{\textbf{Image of study sites used in the sidewalk-shed perception survey.} Each row corresponds to a different location in downtown Manhattan (Locations A–C; A: University Place \& East 12th Street/Broadway, B: West 12th Street \& 7th Avenue South, C: West 12th Street \& Washington Street), and each column shows one of three visual conditions: current sidewalk scaffolding (author-captured), past scaffolding configuration (archival Google Maps imagery), and no scaffolding (baseline Google Maps imagery). The triplet design enables controlled comparison of pedestrian perception and visibility across shed conditions at the same site.}
    \label{fig:site}
\end{figure}

Study sites were selected to systematically vary across these three code-defined features (i.e., clearance height, color, and post spacing), while keeping the survey manageable for participants (Table \ref{tab:site}). Three locations were chosen to ensure sufficient diversity in shed design without requiring participants to evaluate an excessive number of sites. Each location includes three visual scenarios (Figure \ref{fig:site}): (1) a recent image of the sidewalk shed captured by the author, (2) an archival image depicting an earlier shed configuration from Google Maps, and (3) a baseline image showing the site without a shed from Google Maps as well.
Limiting the survey to three sites allows each location to be evaluated in approximately five minutes, resulting in a total survey duration of about fifteen minutes per participant while reducing cognitive load and fatigue. All sites are located in downtown Manhattan, an area characterized by high pedestrian volumes, dense commercial activity, and frequent sidewalk sheds, ensuring that the selected locations are both policy-relevant and representative of real-world urban conditions.

To capture pedestrians' perceptions and navigation decisions, the AI chatbot was designed to integrate visual interaction with guided dialogue. This paper employs an image-based interactive survey that supports direct visual annotation rather than relying solely on text descriptions, thereby better capturing street-level storefront visibility and pedestrian route choices. This design enables participants to indicate what they observe and where they would walk with greater accuracy, while the chatbot guides the interaction through standardized prompts, ensuring consistent task completion across participants.

\begin{figure}[ht]
\centering
\includegraphics[width=\textwidth]{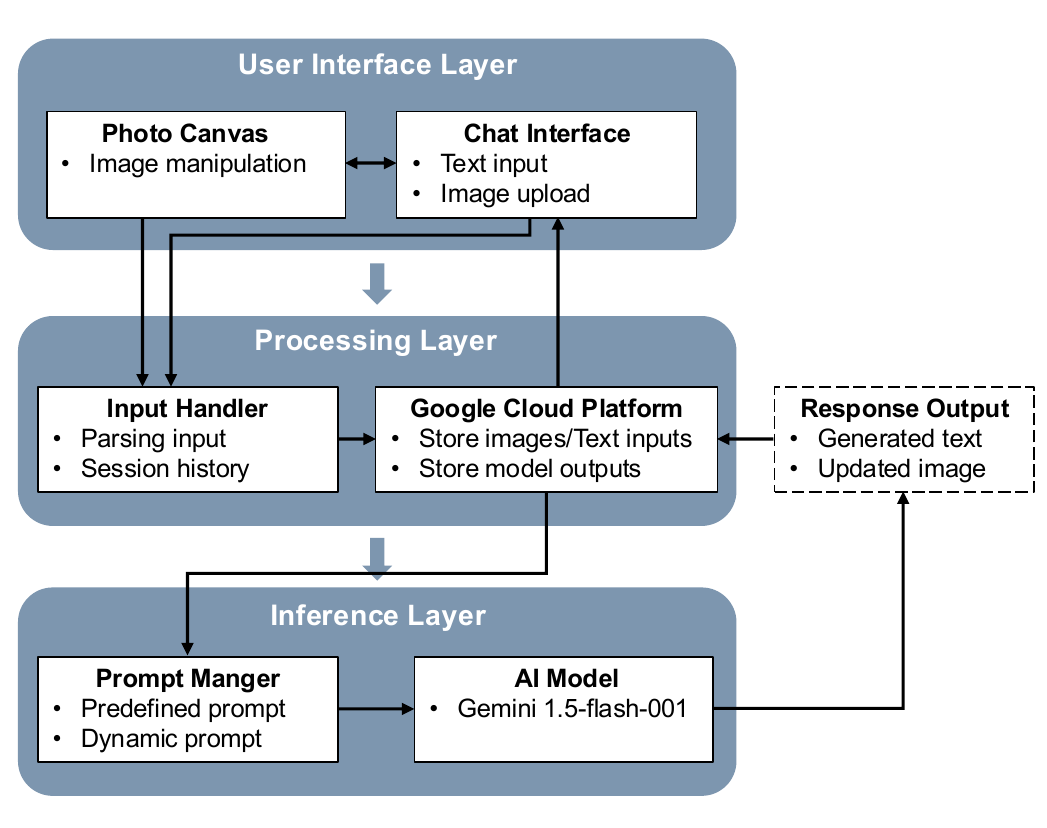}
\caption{\textbf{System architecture of the AI chatbot for pedestrian perception data collection.}
The system consists of three coordinated modules. 
The \textit{User Interface} presents street-level images and allows participants to draw directly on images using a transparent annotation canvas, enabling them to mark storefront entrances, obstructions, and preferred walking paths that are difficult to describe using text alone; a chatbot interface guides users through the tasks via short, adaptive prompts. 
The \textit{Processing Module} receives and organizes user inputs, maintains session context, and securely stores annotated images and dialogue through the Google Cloud Platform. 
The \textit{Inference Module} combines a Prompt Manager with the \textit{Gemini-1.5-flash-001} large language model to interpret user actions and generate consistent, task-oriented responses. 
Together, these components support multi-modal interaction and the collection of spatially explicit pedestrian feedback on sidewalk shed designs.}
\label{fig:architecture}
\end{figure}

The AI chatbot architecture (Figure \ref{fig:architecture}) integrates three primary modules: a `user interface module' a `processing module' and an `inference module'. (1) The `user interface module' comprises two interactive components — an image annotation interface that allows users to draw on images directly via a transparent canvas overlay, and a chatbot interface that facilitates conversational exchanges through text or image inputs. (2) The `processing module' manages user inputs, preserves conversational context through session history, and interfaces seamlessly with the Google Cloud Platform, which securely stores inputs and generated outputs. This backend workflow efficiently coordinates interactions between the `user interface module' and the `inference module'. (3) Within the `inference module', the prompt manager and the \textit{Gemini-1.5-flash-001} large language model collaboratively process textual and contextual data, generating coherent, concise, and contextually relevant responses. Leveraging Gemini 1.5’s rapid inference capability and nuanced understanding, the system ensures effective communication without unnecessarily prolonging interactions.

\begin{figure}
\centering
\includegraphics[width=\textwidth]{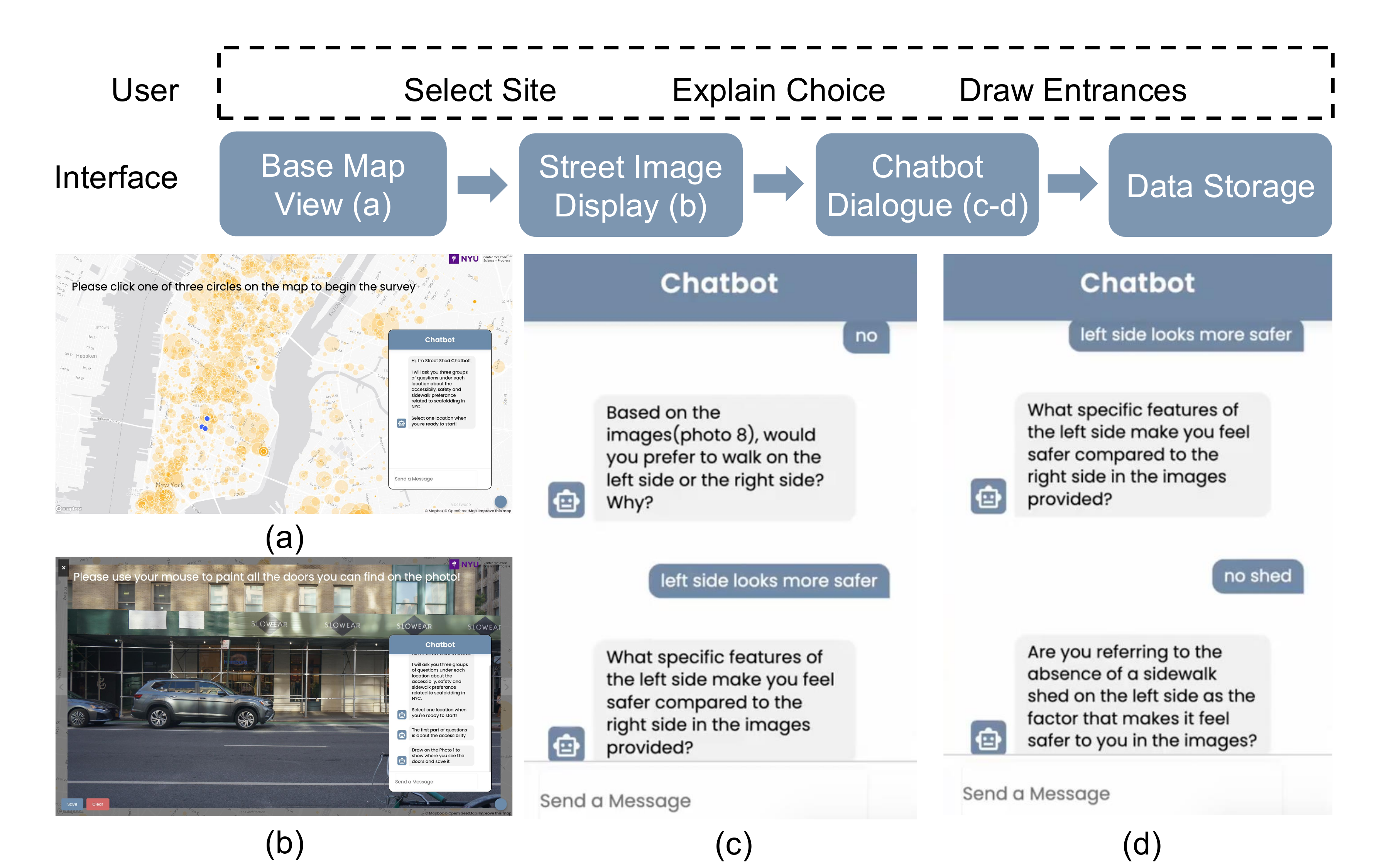}
\caption{\textbf{User interface components of the AI chatbot survey system.}(a) Interactive base map displaying existing sidewalk sheds (orange) and study sites (blue); users click on a site to begin the survey. (b) Photo display area with annotation functionality, allowing users to mark features directly on images. (c–d) chatbot interface guiding participants through text-based survey questions and follow-up prompts based on user responses.}\label{fig:user_interface}
\end{figure}

The user interface was designed with three key components: an interactive base map (Figure \ref{fig:user_interface}(a)), a photo display area (Figure \ref{fig:user_interface}(b)), and a chatbot (Figure \ref{fig:user_interface}(c), Figure \ref{fig:user_interface}(d)), streamlining the survey process and enhancing data quality. (1) The base map displays study sites and existing sidewalk sheds, with blue points representing study sites and orange points indicating existing sheds. (2) As participants complete each site task, the corresponding site marker dynamically updates to gray, providing visual feedback on survey progress. (3) A chatbot, accessible via a button in the bottom-right corner, facilitates participant communication throughout the survey and can be minimized or expanded to maximize the visibility of the map and images as needed. To ensure that participant perceptions are captured with greater precision, the interface incorporates direct annotation tools alongside photographic displays. Eventually, participants could visually mark shop entrances and sidewalk preferences. This combination of visual interaction and dynamic survey flow was designed to promote intuitive engagement while improving the reliability and granularity of the data collected regarding the impacts of sidewalk sheds on pedestrian navigation and accessibility.

\subsection{Experimental Design}

During the survey, participants first selected a study site from an interactive base map. At each selected location, they completed a sequence of tasks, beginning with annotating shop entrances in images under different scaffolding conditions using mouse-based drawing tools. Participants then selected their preferred side of the sidewalk across various design and weather scenarios and provided a rationale for their choices. Upon completing all site tasks, the interface automatically transitioned to a thank-you page. All user-generated inputs, including spatial annotations and text responses, were systematically recorded, enabling the robust collection of spatial and perceptual data on how sidewalk sheds influence pedestrian navigation, accessibility, and urban experiences.

\begin{figure}
\centering
\includegraphics[width=\textwidth]{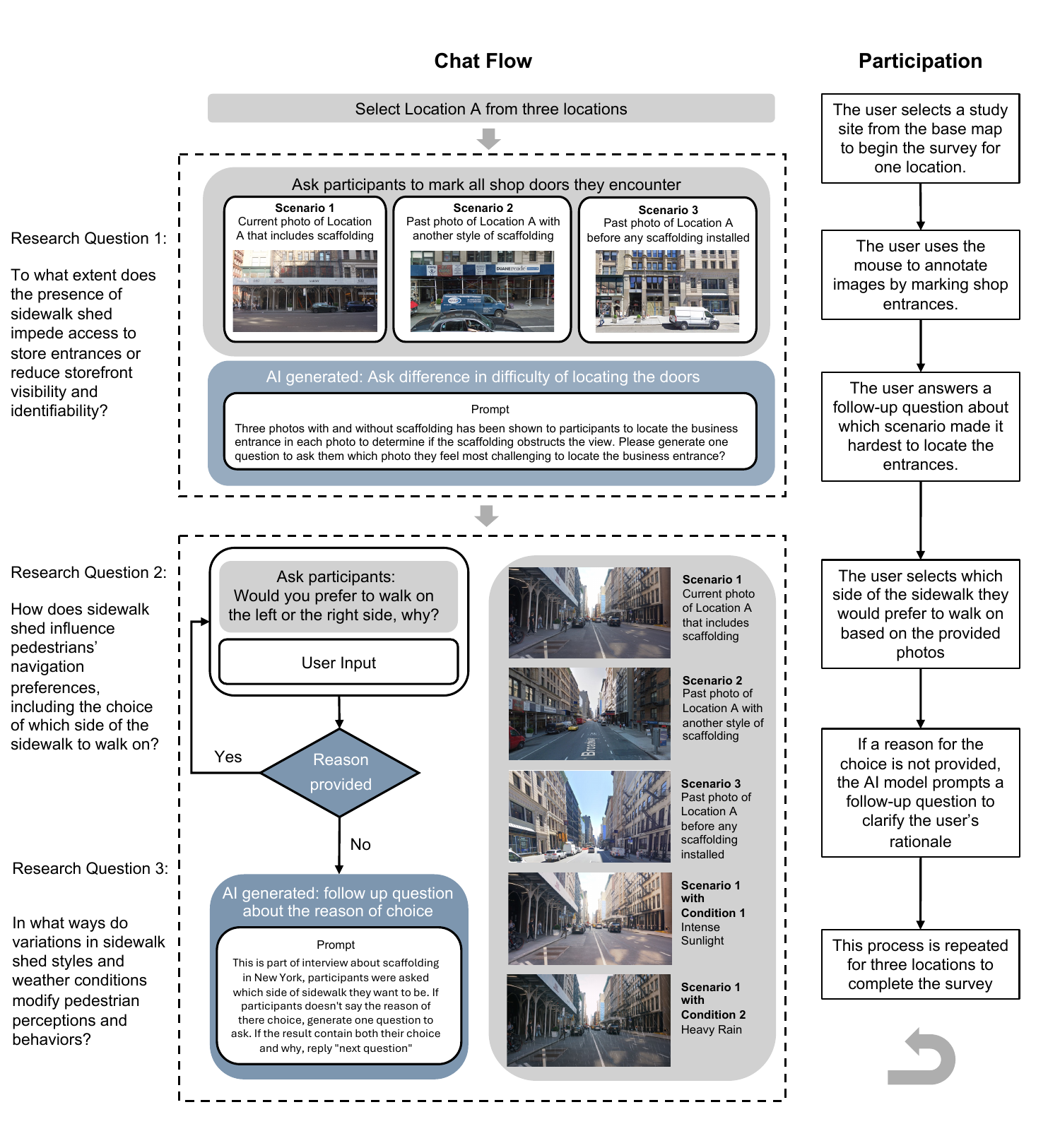}
\caption{\textbf{Chat flow of the AI-assisted survey.} The figure illustrates the end-to-end interaction design used to address three research questions on sidewalk sheds and pedestrian perception. Participants select study locations, annotate storefront entrances under varying scaffolding and environmental conditions, and make sidewalk preference choices. Structured tasks are complemented by AI-generated follow-up prompts that elicit perceived difficulty and reasoning when responses are incomplete. The flow integrates spatial annotations, subjective assessments, and controlled scenario comparisons (design vs. environmental variation), enabling precise measurement of visibility errors, navigation preferences, and contextual influences while maintaining a consistent, participant-centered survey experience.}\label{fig:chat_flow}
\end{figure}

The chat flow (Figure \ref{fig:chat_flow}) was designed to address three research questions, integrating structured tasks with AI-driven dynamic prompts to enhance data quality. To investigate the impact of sidewalk sheds on storefront visibility (Research Question 1), participants were asked to annotate entrances visible in images under varying scaffolding conditions, followed by AI-generated prompts that assessed which scenario most impeded identification. While participants could report the number of entrances identified, textual or numerical responses are insufficient to capture spatial inaccuracies.
Participants may mistakenly identify non-entrance elements (e.g., windows or signage) as doors or fail to mark actual entrances—errors that cannot be distinguished when only aggregate counts are analyzed. 
By requiring participants to annotate the perceived locations of entrances directly on images, the survey collected position-specific data, enabling a more precise and reliable assessment of perceptual errors related to visibility under sidewalk sheds. Moreover, the AI-generated follow-up questions were designed to quantify the perceived difficulty of the identification task, recognizing that even successful entrance recognition may entail differing levels of cognitive effort, including variations in the time spent and attentional demands. By capturing both objective performance and subjective difficulty, the chat flow enabled a more comprehensive evaluation of the visual disruptions caused by sidewalk sheds.

In the second stage, participants were prompted to select a preferred side of the sidewalk and justify their choices, thereby enabling the collection of behavioral preferences, which is necessary for understanding how scaffolding influences pedestrian navigation decisions (Research Question 2). In instances where responses lacked sufficient elaboration, the AI module dynamically generated clarification prompts to ensure the consistent capture of participants’ underlying reasoning. To examine the influence of environmental factors on behavior (Research Question 3), simulated weather scenarios were seamlessly integrated into the sidewalk choice tasks, allowing for controlled comparisons without increasing participants' cognitive burden. The five image sets were structured to support two types of paired experimental comparisons: one varying environmental conditions while holding physical design constant (different weather conditions, same scenario) and the other varying design features while maintaining consistent environmental conditions (same weather condition, different scenarios). This dual structure enabled within-subject comparisons to disentangle the distinct effects of environmental and design variables.

All user-generated inputs, including spatial annotations and textual responses, were systematically recorded and flagged based on whether they addressed predefined prompts or dynamically generated follow-up questions, thereby preserving both the rigor of scripted evaluation and the flexibility of adaptive inquiry. By restricting the AI model’s context window to the three most recent user interactions, the system balanced the need for conversational continuity with computational efficiency. Through this integrated design, the chat flow facilitated robust, high-quality data collection across all research dimensions while maintaining a coherent and participant-centered survey experience.

\subsection{Participation and Data Collection}

The survey was conducted online between July 24th and October 30th, 2025, using the Qualtrics platform (\url{https://www.qualtrics.com}) to distribute the questionnaire. Participants were invited through convenience sampling (e.g., university mailing lists and social groups), utilizing both digital and interpersonal channels. Eligible participants were individuals who had previously visited New York City. A total of 46 responses were collected, of which 25 complete submissions were retained for analysis after excluding incomplete entries. The final analytic sample recorded an average survey completion time of 15 minutes and 7 seconds.

\subsection{Data Analysis}

To address Research Question 1, which asks whether sidewalk sheds impede storefront visibility and perceptual discrimination, we analyzed image-annotation data generated by participants marking store entrances. These data are spatial, repeated, and hierarchical, with multiple images evaluated by the same participants and multiple observations nested within locations.

Given this structure, mixed-effects models were fitted in Python using \texttt{statsmodels} (v0.14.2) to estimate the association between sidewalk-shed exposure and perceptual sensitivity while accounting for repeated observations within participants and locations:
\begin{equation}
d'_{ij} = \beta_0 + \beta_1\,\textit{S}_{ij} + u_{\text{p}[i]} + v_{\text{l}[j]} + \varepsilon_{ij},
\end{equation}
\begin{equation}
u_{\text{p}[i]} \sim \mathcal{N}(0,\,\sigma^2_{\text{p}}), 
\qquad
v_{\text{l}[j]} \sim \mathcal{N}(0,\,\sigma^2_{\text{l}}),
\qquad
\varepsilon_{ij} \sim \mathcal{N}(0,\,\sigma^2).
\end{equation}
where $\textit{S}_{ij}$ indicates whether the image contains a sidewalk shed (0 = no shed, 1 = shed); $u_{\text{p}[i]}$ is a participant-specific random intercept; $v_{\text{l}[j]}$ is a location-specific random effect;$\varepsilon_{i j}$ is the residual error term, and $d'_{ij}$ denotes the signal-detection sensitivity ($d'$) for participant $i$ on image $j$. 
Sensitivity ($d^\prime$) was selected as the outcome variable because it jointly captures hit rates and false-alarm rates, providing a robust measure of discrimination that is widely used in vision and human-perception research.

To derive $d'$, a grid-based evaluation was employed to provide spatial tolerance and mitigate minor misalignment between participant strokes and ground-truth entrances, consistent with established practices in object-detection benchmarking \citep{everingham2010pascal}.
Each image was processed using the \texttt{Pillow} (v12.0.0) library and divided into 50$\times$50-pixel grids, corresponding to the stroke width used in participant annotations. Ground-truth entrance regions were manually delineated based on visual inspection of the original images. 
For each grid cell, annotation outcomes were classified as true positive (TP), false positive (FP), false negative (FN), or true negative (TN) following standard binary-evaluation definitions \citep{sokolova2009systematic}. 
A log-linear correction was applied to avoid infinite values in the $z$ -transform \citep{hautus1995corrections}, whereby corrected hit and false-alarm rates were computed as
\begin{equation}
H = \frac{\mathrm{TP} + 0.5}{\mathrm{TP} + \mathrm{FN} + 1},
\qquad
F = \frac{\mathrm{FP} + 0.5}{\mathrm{FP} + \mathrm{TN} + 1}.
\end{equation}
Sensitivity was then calculated as
\begin{equation}
d' = \Phi^{-1}(H) - \Phi^{-1}(F),
\end{equation}
where $\Phi^{-1}$ denotes the inverse normal cumulative distribution function. 
Participant–location cases with sensitivity ($d^\prime$) below 1, reflecting unreliable discrimination performance, were excluded. To support interpretation, heatmaps of aggregated annotations were generated to visualize spatial patterns of attention and misidentification under shed and no-shed conditions. These visualizations complement the statistical models by illustrating how sidewalk sheds alter the spatial distribution of perceived storefront locations.

To address Research Questions 2 and 3, which examine how sidewalk sheds, environmental conditions, and design features influence pedestrians’ choice of walking side, we analyzed participants’ binary route-selection responses. Because the outcome is dichotomous (shed side vs. non-shed side), binomial logistic regression was used.

The general form is as follows:
\begin{equation}
\Pr(Y_i = 1 \mid \mathbf{X}_i)
= \text{logit}^{-1}(\eta_i)
= \frac{1}{1 + \exp(-\eta_i)},
\end{equation}
where $Y_i = 1$ indicates that participant $i$ chose the shed side and $Y_i = 0$ otherwise. The linear predictor is expressed as
\begin{equation}
\eta_i = \beta_0 + \boldsymbol{\beta}^{\top}\mathbf{X}_i,
\end{equation}
with $\mathbf{X}_i$ containing binary indicator variables. 
Two sets of models were estimated to reflect the structure of the research questions.
First, to evaluate the role of environmental conditions, the predictors included the dummy variables
$X_{i,\text{Normal}}$, $X_{i,\text{Sun}}$, and $X_{i,\text{Rain}}$, indicating whether a trial was conducted under \textit{Normal Weather}, \textit{Harsh Sunlight}, or \textit{Heavy Rain}, respectively.
Second, to isolate the effects of shed design attributes, predictors included binary factors indicating whether the shed was green in color ($X_{i,\text{green}}$), had more supporting columns ($X_{i,\text{column}}$), or had a lower height ($X_{i,\text{low}}$). Both model use \textit{No Shed} as the reference category. Coefficients, therefore, represent the change in log-odds of selecting the shed side associated with the presence versus absence of each attribute.
Model adequacy for both models was evaluated using McFadden’s pseudo $R^{2}$, defined as
\begin{equation}
R^2_{\text{McF}} 
= 1 - \frac{\ell(\hat{\boldsymbol{\beta}})}{\ell(\hat{\boldsymbol{\beta}}_{0})},
\label{eq:mcfadden}
\end{equation}
where $\ell(\hat{\boldsymbol{\beta}}_{0})$ is the log-likelihood of the intercept-only model.
This metric provides an interpretable measure of explanatory power for discrete-choice models.

\section{Results}
\subsection{Analysis of Entrance Visibility}

\begin{figure}
\centering
\includegraphics[width=\textwidth]{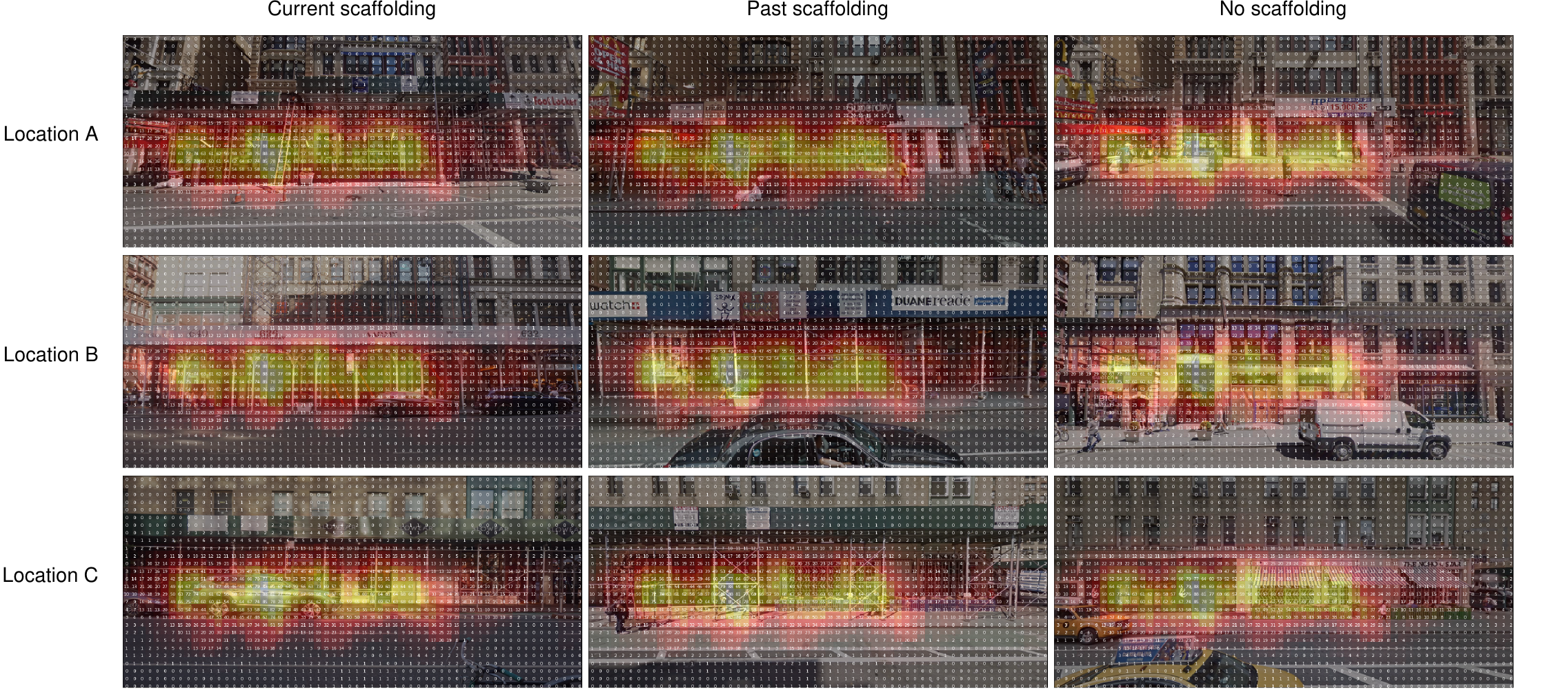}
\caption{
\textbf{Heatmaps of participant annotation density across sidewalk shed conditions.}
Each image is $2988\times1380$ pixels and subdivided into $50\times50$-pixel grids; numbers indicate the count of annotations within each grid cell.
Rows correspond to study locations (A–C), and columns show the three visual scenarios: current sidewalk shed, past sidewalk shed, and no shed.
Warmer colors indicate higher concentrations of participant annotations identifying storefront entrances.
Across all locations, annotations are more spatially dispersed and less concentrated under current and past shed conditions, whereas the no-shed condition shows tighter clustering around storefront entrances.
This pattern suggests that the presence of sidewalk sheds reduces the visual salience and visibility of storefront entrances.
}
\label{fig:heatmap}
\end{figure}

To address Research Question 1, this study analyzes participants’ storefront-entrance annotations using spatial heatmaps, signal-detection sensitivity measures ($d^\prime$), and mixed-effects modeling.
Across all three locations, participants consistently demonstrated reduced entrance-recognition performance when sidewalk sheds were present. 
Heatmap visualizations (Figure \ref{fig:heatmap}) illustrate this effect spatially. Under both current and past shed conditions, participant annotations are more diffuse and spatially inconsistent. In contrast, under the no-shed condition, annotations cluster tightly around the actual doorway areas.

\begin{figure}
\centering
\includegraphics[width=0.8\textwidth]{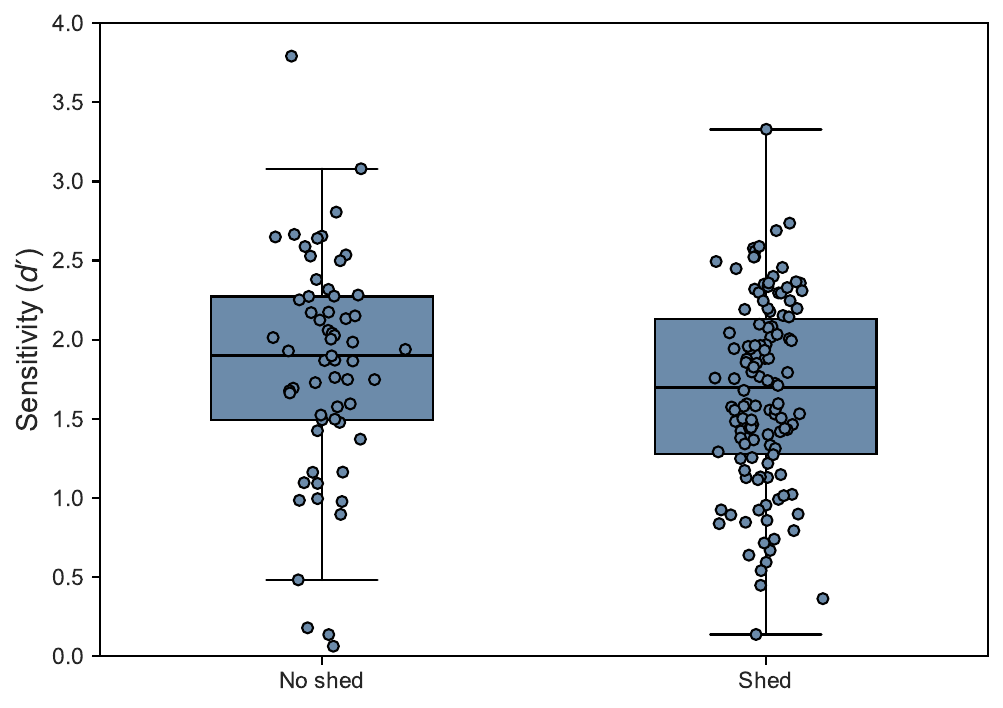}
\caption{\textbf{Sensitivity ($d^\prime$) under shed and no-shed conditions.} 
Each point represents a participant–location observation (after excluding cases with $d^\prime < 1$), and values reflect signal-detection sensitivity to identify designated entrance regions. Higher $d^\prime$ indicates better perceptual discrimination. Error bars show 95\% confidence intervals estimated from the mixed-effects model.}
\label{fig:img-result}
\end{figure}
 
These visual patterns are corroborated by the sensitivity analysis (Figure~\ref{fig:img-result}). Participant–location sensitivity scores ($d^\prime$), which quantify the ability to differentiate true entrances from non-entrances, are systematically lower when a sidewalk shed is present. Although absolute sensitivity levels vary by location due to differences in façade configuration and lighting, the overall pattern indicates lower $d^\prime$ values under shed conditions compared to the no-shed baseline.

\begin{table}[ht]
\centering
\caption{\textbf{Mixed-effects model estimating the association between sidewalk-shed exposure and perceptual sensitivity ($d'$).} 
The analysis included 25 participants and 183 participant-level observations across three locations.}
\begin{tabular}{lccc}
\toprule
\textbf{Parameter} & \textbf{Coefficient} & \textbf{$p$-value} & \textbf{95\% CI} \\
\midrule
Intercept ($\beta_0$) & $1.811$ & $<0.001$ & $[1.613,\,2.008]$ \\
Shed effect ($\beta_1$) & $-0.161$ & $0.021$ & $[-0.298,\,-0.024]$ \\
\midrule
\multicolumn{4}{l}{\textbf{Random effects}} \\
Participant variance ($\sigma^2_{\text{p}}$) & $0.129$ & & \\
Location variance ($\sigma^2_{\text{l}}$)    & $0.073$ & & \\
Residual variance ($\sigma^2$)               & $0.199$ & & \\
\midrule
\multicolumn{4}{l}{\textbf{Model fit}} \\
\multicolumn{4}{l}{\quad Pseudo $R^{2}$ = $0.614$} \\
\multicolumn{4}{l}{\quad Marginal $R^{2}$ = $0.372$ \quad Conditional $R^{2}$ = $0.542$} \\
\multicolumn{4}{l}{\quad Log-likelihood = $-148.740$} \\
\bottomrule
\end{tabular}
\label{tab:img}
\end{table}

Mixed-effects modeling also shows that exposure to sidewalk-sheds is associated with a statistically significant decrease in perceptual sensitivity ($\beta_1 = -0.161$, $p = 0.021$) in Table \ref{tab:img}, indicating that sheds impair the visual recognition of entrances. Together, these results demonstrate that the presence of sidewalk sheds reduces the clarity of entrance cues and undermines users’ ability to visually identify building access points.

\subsection{Human Preference under Weather Conditions }

\begin{figure}
\centering
\includegraphics[width=\textwidth]{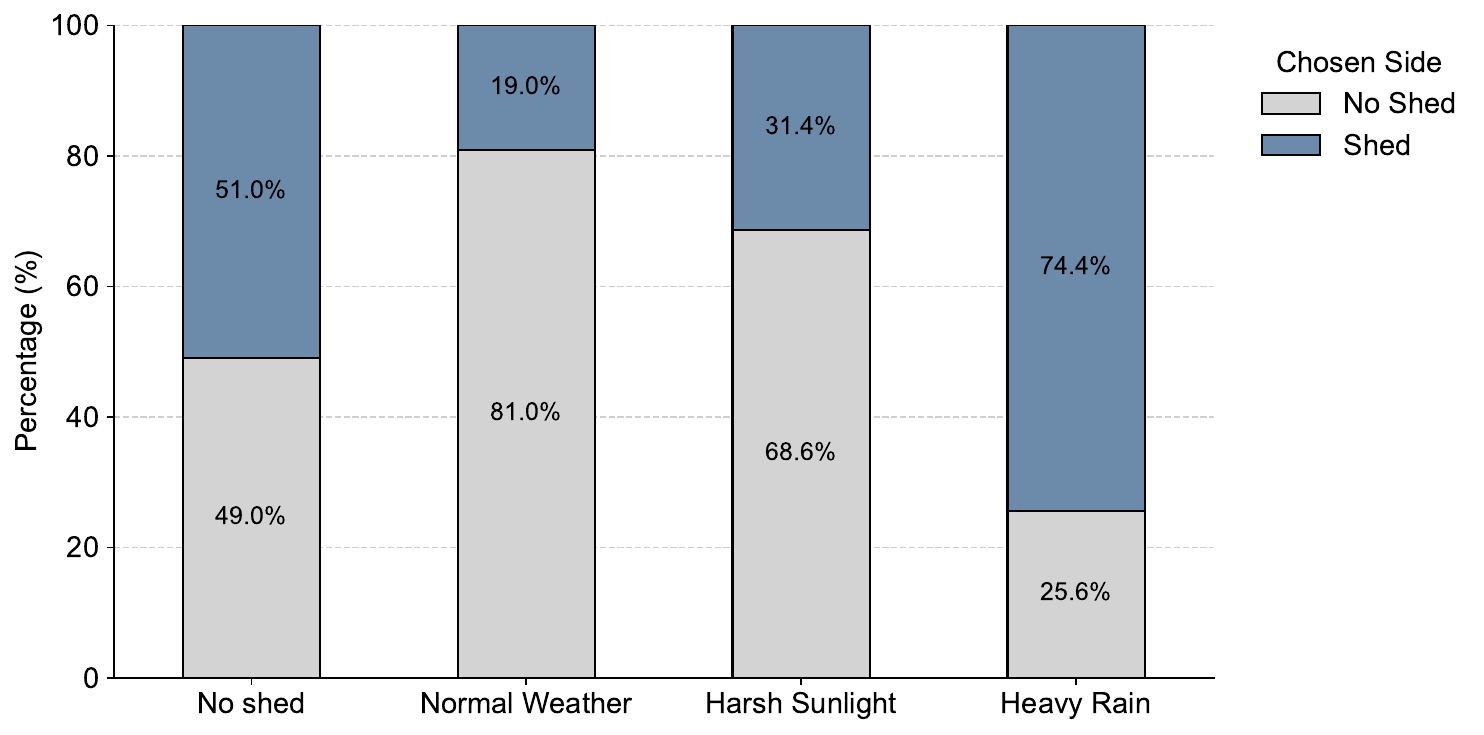}
\caption{\textbf{ Proportion of participants choosing the shed side under four weather conditions.} The figure illustrates a clear weather-dependent shift in route preference: participants strongly avoided the shed side in normal weather, showed modestly increased use under harsh sunlight, and overwhelmingly preferred the shed during heavy rain. These descriptive patterns highlight the functional value of shelter when environmental conditions deteriorate.}\label{fig:weather-result}
\end{figure}

\begin{table}[ht]
\centering
\caption{\textbf{Logistic regression estimates for the effect of weather conditions on shed-side choice.} Coefficients are expressed relative to the reference category (No shed condition). The model indicates significant avoidance of the shed side in normal weather and a strong, statistically significant increase in shed-side selection during heavy rain, demonstrating a substantial weather-driven shift in pedestrian routing behavior.}
\begin{tabular}{lccc}
\toprule
\textbf{Parameter} & \textbf{Coefficient} & \textbf{$p$-value} & \textbf{95\% CI} \\
\midrule
Intercept ($\beta_0$)                 & $-0.615$ & $0.002$   & $[-1.000,\,-0.231]$ \\
Normal Weather ($\beta_{\text{Normal}}$)      & $-0.794$ & $0.042$   & $[-1.557,\,-0.030]$ \\
Harsh Sunlight ($\beta_{\text{Sun}}$) & $0.022$  & $0.947$   & $[-0.635,\,0.679]$  \\
Heavy Rain ($\beta_{\text{Rain}}$)    & $1.737$  & $<0.001$  & $[1.022,\,2.453]$   \\
\midrule
\multicolumn{4}{l}{\textbf{Model Fit}} \\
\multicolumn{4}{l}{\quad Log-likelihood = $-171.80$} \\
\multicolumn{4}{l}{\quad McFadden's pseudo $R^{2}$ = $0.109$}\\
\bottomrule
\end{tabular}
\label{tab:weather}
\end{table}

Weather conditions substantially influenced pedestrians’ likelihood of selecting sidewalks with or without scaffolding (Figure~\ref{fig:weather-result}). During normal weather, the majority of participants (81\%) preferred the unsheltered side, indicating an overall tendency to avoid the shed when ambient conditions were comfortable. However, the tendency to avoid the shed decreased as the weather became more adverse. Under harsh sunlight, shed-side use increased to 31.4\%. Moreover, this pattern reversed in heavy rain: 74.4\% of participants chose the shed side, indicating a strong behavioral shift toward shelter when protection from the elements became salient. The regression result (Table \ref{tab:weather}) is consistent with the trends. Relative to the baseline no-shed condition, the likelihood of selecting the shed side was significantly lower in normal weather ($\beta$ = –0.794, $p$ = 0.042), confirming strong avoidance under fair conditions. The effect under harsh sunlight conditions was not statistically significant ($\beta$ = 0.022, $p$ = 0.947), indicating that increased brightness did not affect behavior across participants. In contrast, heavy rain produced a pronounced and statistically significant increase in the probability of choosing the shed side ($\beta$ = 1.737, $p$ < 0.001), demonstrating that severe weather sharply heightened the demand for shelter.

\subsection{Factors of Shed Design Styles}

\begin{figure}
\centering
\includegraphics[width=\textwidth]{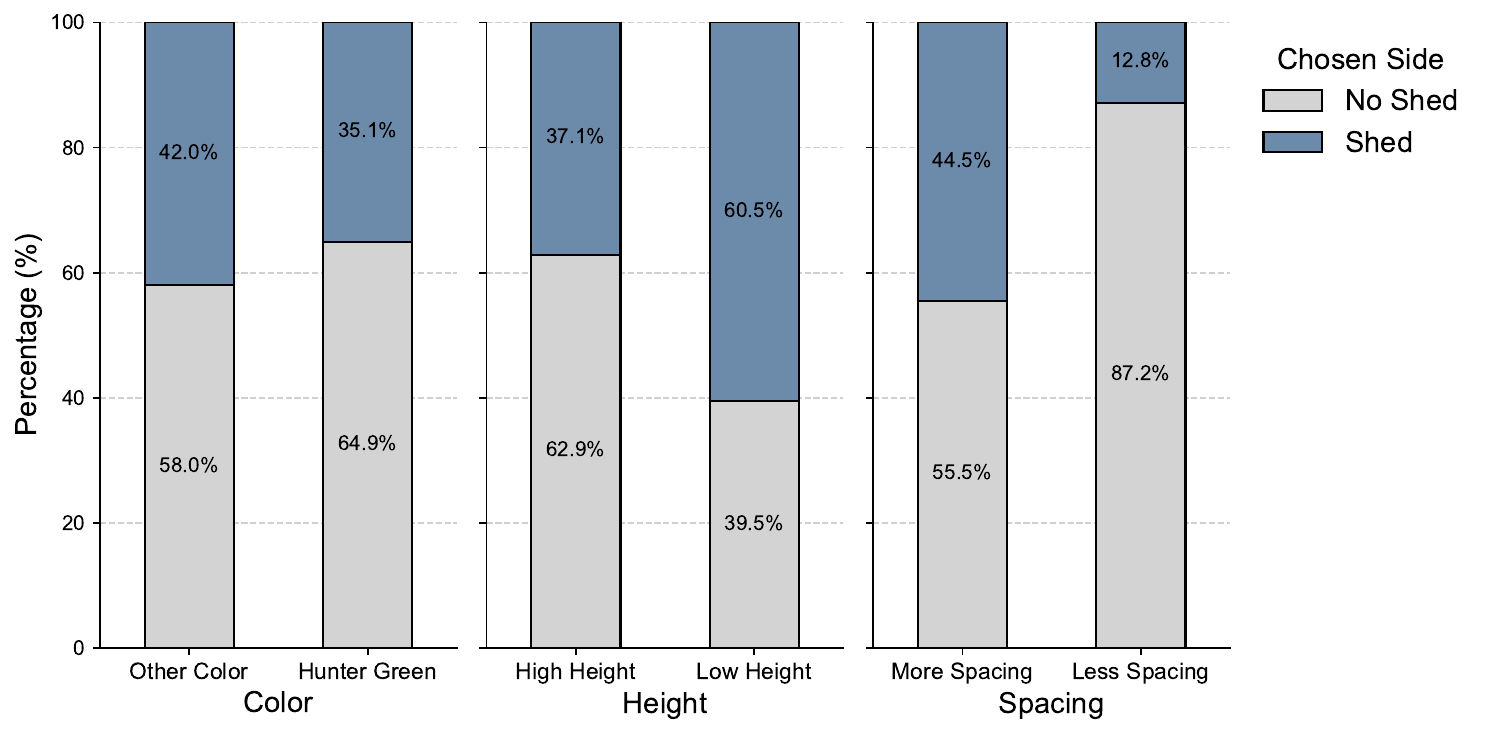}
\caption{\textbf{Participant choices between the shed and non-shed sides under different shed-design attributes.} Bars represent the percentage of participants selecting the shed side for each design condition. Colors in the legend correspond to the attribute being varied: (a) shed color, (b) shed height, and (c) post spacing. Each bar reflects one design variant within the corresponding attribute category.}
\label{fig:design-result}
\end{figure}

\begin{table}[ht]
\centering
\caption{\textbf{Estimated coefficients from logistic regression models evaluating shed-design features.} Predictors include color, height, and post spacing, with results presented as coefficients, standard errors, z-values, and p-values relative to their reference categories.}
\begin{tabular}{lccc}
\toprule
\textbf{Parameter} & \textbf{Coefficient} & \textbf{$p$-value} & \textbf{95\% CI} \\
\midrule
Intercept ($\beta_0$)                        & $-0.286$ & $0.057$  & $[-0.579,\, 0.008]$ \\
Green color ($\beta_{\text{green}}$)         & $-0.303$ & $0.329$  & $[-0.912,\, 0.305]$ \\
Less spacing ($\beta_{\text{column}}$)       & $-1.500$ & $0.003$  & $[-2.490,\,-0.510]$ \\
Low height ($\beta_{\text{low}}$)            & $0.867$  & $0.021$  & $[0.128,\, 1.606]$ \\
\midrule
\multicolumn{4}{l}{\textbf{Model Fit}} \\
\multicolumn{4}{l}{\quad Log-likelihood = $-181.87$} \\
\multicolumn{4}{l}{\quad McFadden's pseudo $R^{2}$ = $0.056$} \\
\bottomrule
\end{tabular}
\label{tab:design}
\end{table}

Participants’ willingness to walk under a sidewalk shed may primarily depend on how open, navigable, and visually permeable the structure appears. Figure \ref{fig:design-result} illustrates how pedestrian preferences varied across three shed-design attributes: color, height, and post spacing. Differences across color conditions were modest, with shed-side selections remaining relatively balanced, regardless of whether the shed was hunter green or another color. Height produced a more noticeable contrast: lower sheds were chosen by 60.5\% of participants, whereas only 37.1\% selected the shed side when the structure was taller, suggesting that height influenced the immediate perception of enclosure. Post spacing yielded the clearest differences. When posts were spaced farther apart, 44.5\% of participants chose the shed side, compared with only 12.8\% when the spacing was narrower. This pattern indicates that pedestrians responded sensitively to changes in spatial openness and maneuvering space beneath the structure. The logistic regression results (Table \ref{tab:design}) complement these patterns. By identifying design attributes that predict street-side selection, we find that narrower post spacing reduces the likelihood of pedestrians choosing the side with scaffolding ($\beta$ = –1.500, $p$ = 0.003). Low shed height showed a significant positive effect ($\beta$ = 0.867, $p$ = 0.021), indicating that lower structures were more likely to be selected, even after accounting for other variables. In contrast, color had no statistically significant impact on pedestrian choice ($\beta$ = –0.303, $p$ = 0.329), reflecting the minimal variation observed in Figure \ref{fig:design-result}.

\section{Discussion}

This paper evaluates the impacts of sidewalk sheds on pedestrian perception and choice using an AI-based chatbot survey. The results demonstrate that scaffolding reduces entrance visibility and that shed design features significantly shape sidewalk choice behavior. This paper shows the feasibility of using conversational AI chatbots to collect public opinions in urban management.

The analysis of entrance visibility indicates a `perceptual cost' associated with sidewalk sheds. Perceptual cost refers to the cognitive effort, time, or difficulty experienced when processing sensory information, making decisions, or adapting to changes \citep{suzuki2009perceptual}. Sidewalk sheds obstruct pedestrians’ visual access to building entrances, thereby reducing the reliability with which they can identify points of entry. The heatmaps demonstrate how scaffolding disrupts the salience of architectural cues by introducing visual clutter and partial occlusion, consistent with prior evidence that extraneous structures weaken environmental legibility \citep{Florio2024VisualComplexity}. This perceptual interference aligns with broader concerns about the economic consequences of prolonged scaffolding installations \citep{kang2016high,Mayor2024,Menkes2003}. When storefront entrances are visually fragmented or partially hidden, potential customers may overlook businesses entirely or misinterpret them as closed or inaccessible—an effect previously linked to reduced spontaneous visits and diminished window-display effectiveness in urban retail environments \citep{mower2012exterior}. Thus, the reduced entrance-recognition accuracy observed in both behavioral heatmaps and sensitivity modeling likely reflects not only a perceptual barrier for pedestrians but also a mechanism through which sidewalk sheds can negatively impact street-level commercial activity.

Results for different weather conditions show that sidewalk sheds influence pedestrian navigation in a context-dependent manner. Under typical conditions, sheds tend to reduce walking comfort and perceived environmental quality, leading to consistent avoidance. However, when environmental stress increases (e.g., heavy rain), the functional value of shelter outweighs these drawbacks, resulting in a marked shift in preference toward the shed side.
This behavioral reversal is consistent with prior studies showing that pedestrians actively adapt their routes to reduce weather-related discomfort by seeking overhead protection or building edges \citep{Dong2022WeatherPedestrians,BagheriMovahhed2020RainPedestrians}. These findings suggest that sidewalk sheds simultaneously function as perceptual barriers and protective infrastructure, depending on the environmental context.
This duality underscores the importance of considering the environmental context when evaluating temporary streetscape structures: while sidewalk sheds may diminish the pedestrian experience, they become essential protective infrastructure during extreme weather. Design strategies that enhance openness, visibility, or light quality may help reduce pedestrian's perceived burden under typical conditions while preserving their value as shelter during severe weather.

These findings regarding design styles further indicate that pedestrians respond more strongly to spatial and structural characteristics than to surface aesthetics.
Wider post spacing appears to improve visual permeability and movement comfort, making the covered route feel less constrained \citep{Zhou2025UrbanVisualAttractiveness}. The preference for lower sheds, though initially unexpected, may reflect a perceptual trade-off between internal and external viewpoints: while higher sheds provide more headroom, they also introduce larger, more visually dominant elements into the streetscape, potentially making the structure appear more imposing from a distance. The negligible influence of color suggests that aesthetic variation alone is insufficient to shift behavior in functional, decision-oriented contexts. From an urban-design perspective, these results underscore the importance of designing sidewalk sheds that balance protection with spatial openness, clear sightlines, and minimized visual mass—qualities that can enhance pedestrian comfort and maintain walkability even within construction-dominated streetscapes.

This study has several limitations. First, the analysis is limited to downtown Manhattan, a dense and highly regulated urban environment; thus, the findings may not generalize to cities with different street widths, building typologies, or construction practices. Second, the sample size and the relatively short data collection window may introduce sampling and temporal biases. Third, the study focuses on a limited set of design attributes defined by current building codes and does not account for other potentially influential factors, such as lighting conditions, nighttime visibility, or pedestrian crowding. Finally, over half of the participants did not complete the survey, likely due to its length, repetitive task structure, and technical and accessibility constraints. In particular, the survey was only accessible via laptop and required redirection to an external website, introducing additional friction during participation and contributing to participant attrition. This challenge reflects a common limitation of digital platforms for public data collection: motivating initial participation and minimizing access barriers remain critical obstacles \citep{newman2021data}.

Future research should extend this framework by broadening the geographic scope, recruiting more demographically diverse participants, and examining a wider range of design features such as lighting conditions, lateral clearance, aesthetic treatments, and night-time visibility. Integrating computer vision–based scene analysis or immersive virtual reality environments may further enhance the precision of perceptual measurements. Ultimately, scaling this methodology can contribute to more comprehensive, evidence-informed policies for temporary construction infrastructure, advancing urban design practices and pedestrian-centered planning.

The findings highlight several design and regulatory implications. Sidewalk-shed regulations should be updated to explicitly address visual permeability and pedestrian comfort, in addition to safety requirements. Increasing the minimum post spacing and adopting flexible height standards tailored to the street context could mitigate the negative impacts on storefront visibility while preserving the protective function of sheds. Such design-informed regulations would allow cities to maintain safety during construction without disproportionately undermining pedestrian navigation and street-level economic activity.

\section{Conclusion}

In summary, this study presents a novel framework for evaluating sidewalk shed design by integrating an AI-driven chatbot survey with image-based interaction and annotation tools. This approach addresses a critical gap in previous research by generating quantitative and context-rich evidence on how specific physical design elements, such as post spacing, height, and visual openness, influence pedestrian perception and behavior.
The results reveal that sidewalk sheds entail a fundamental trade-off between safety provision and street-level experience, with design choices mediating pedestrian comfort, visibility, and engagement.
These insights offer actionable guidance for policymakers, regulators, and urban designers seeking to balance safety requirements with high-quality pedestrian environments.

\section*{Acknowledgement}

The authors did not receive any external funding for this research.

\section*{Acknowledgment of Generative Al}

Paper writing - The authors used a generative artificial intelligence (Al) tool to assist with language editing and minor formatting improvements during manuscript preparation. The AI system was not used for content generation, interpretation, or decision-making. All final content was reviewed and approved by the authors, who take full responsibility for the integrity of the work.

Coding - A generative AI-assisted coding tool (ChatGPT) was used to assist with troubleshooting error messages during code development. The AI system was not used to make analytical or methodological decisions. All final code was reviewed, tested, and approved by the authors.

\section*{Disclosure statement}

The authors report that there are no competing interests to declare

\subsection*{Ethics or approval statement}
This study was reviewed and approved by the Institutional Review Board (IRB); Title: Using AI Chatbots to Study on Human Opinion of Urban Space (IRB-FY2025-10241). We confirmed that informed consent was obtained from all participants, and all research was conducted in accordance with relevant guidelines and regulations.

\section*{Data Availability Statement}

The participants in this study did not provide written consent for their data to be shared publicly; therefore, due to the sensitive nature of the research, supporting data is not available.

\bibliographystyle{apacite}
\bibliography{bib}

\end{document}